\documentclass[%
 reprint,
 amsmath,amssymb,
 aps,prl
]{revtex4-2}

\usepackage{graphicx}
\usepackage{dcolumn}
\usepackage{bm}
\usepackage{amsmath}
\usepackage[caption=false,subrefformat=parens,labelformat=parens]{subfig}
\usepackage{float}
\usepackage[utf8]{inputenc}
\usepackage[T1]{fontenc}
\usepackage{url}
\usepackage{verbatim}

\begin{document}

\preprint{APS/123-QED}

\title{Dirac Wire}

\author{Farhad Khosravi$^{1,2}$, Todd Van Mechelen$^{2}$, and Zubin Jacob$^{1,2}$}
 \email{zjacob@purdue.edu}
 \affiliation{$^{1}$Department of Electrical and Computer Engineering, University of Alberta, Edmonton, Alberta T6G 1H9, Canada}

 \affiliation{$^{2}$Birck Nanotechnology Center and Purdue Quantum Center, Purdue University, West Lafayette, IN 47906, USA}%

\date{\today}

\begin{abstract}
The interplay of photon spin and orbital angular momentum (OAM) in the optical fiber (1D waveguide) has recently risen to the forefront of quantum nanophotonics. Here, we introduce the fermionic dual of the optical fiber, the Dirac wire, which exhibits unique electronic spin and OAM properties arising from confined solutions of the Dirac equation. The Dirac wires analyzed here represent cylindrical generalizations of the Jackiw-Rebbi domain wall and the minimal topological insulator, which are of significant interest in spintronics. We show the unique longitudinal spin arising from electrons confined to propagation in a wire, an effect which is fundamentally prohibited in planar geometries.  Our work sheds light on the universal spatial dynamics of electron spin in confined geometries and the duality between electronic and photonic spin.
\end{abstract}

\maketitle



\textit{Introduction.---}
Confined solutions of Maxwell's equations exhibit unique phenomena such as transverse photon spin and universal spin-momentum locking \cite{sague2007cold,petersen2014chiral,van2016universal,kalhor2016universal,sollner2015deterministic,rodriguez2013near}. These effects do not occur in conventional circularly polarized propagating plane waves where the spin is always directed longitudinally along the momentum vector \cite{berry1998paraxial}. One striking example is an optical fiber where Zeeman transitions in a cold atom shows spin-dependent directional photon transport \cite{petersen2014chiral,sague2007cold}. The goal of this paper is to introduce the concept of Dirac waveguides and understand the intriguing spin characteristics of confined electronic waves. Our work is motivated by the Dirac-Maxwell correspondence \cite{barnett2014optical,bialynicki1996v} which studies the relativistic parallels between photons and electrons.

\begin{figure}[t!]
    \centering
        \subfloat[\label{Fig:Schematic}]{
        \includegraphics[width=0.98\linewidth]{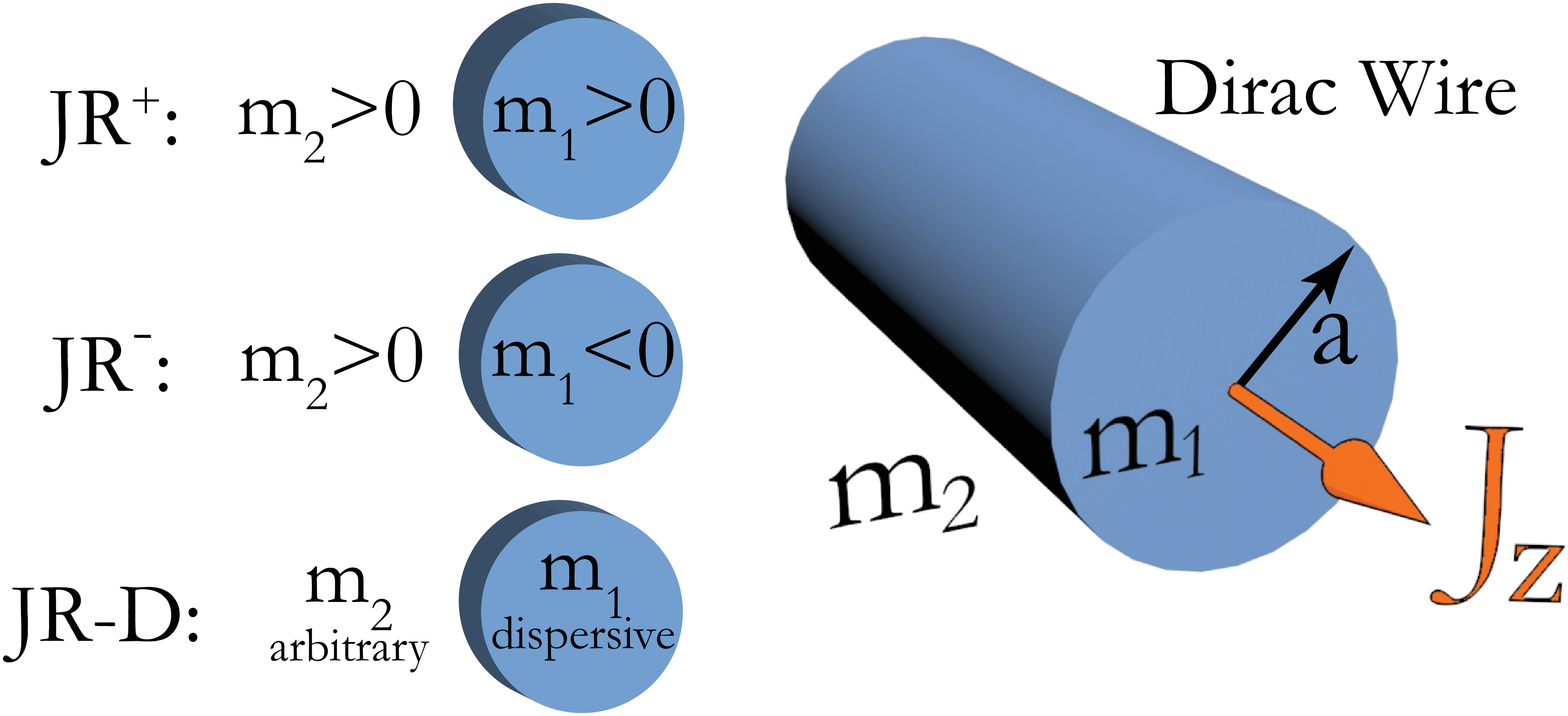}}\par
        \subfloat[\label{Fig:Field_profiles}]{
        \includegraphics[width=0.98\linewidth]{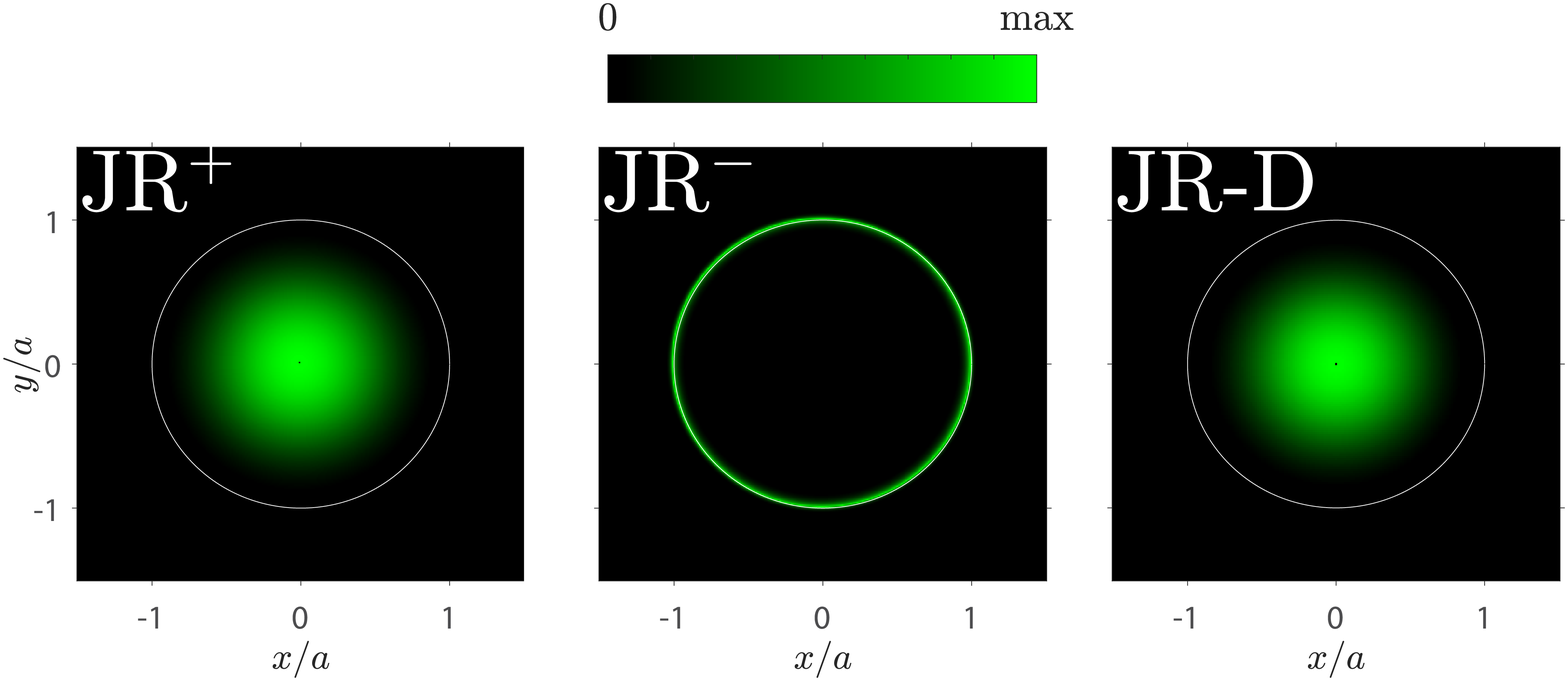}}
    \caption{Schematic of the Dirac wires. (a) The three Jackiw-Rebbi (JR) type domains considered here are JR$^+$ with electron mass inside ($m_1$) and outside ($m_2$) the wire both positive, JR$^-$ with positive mass inside and negative mass outside, and JR-D with a dispersive electronic mass inside [Eq.~(\ref{Eq:Dispersive_mass})] and an arbitrary mass outside. JR-D corresponds to the minimal topological insulator. (b) Distribution of the probability density, $\psi^\dagger \psi$, for the three problems. The fields are normalized such that $\int \psi^\dagger \psi = 1$ when integrated over the entire cross section. Notice that the probability amplitude of the JR$^-$ state is localized around the perimeter of the wire $\rho = a$. Also, in the case of the JR-D problem, the wave function is identically zero at the boundary and outside the wire $\psi(\rho\geq a) = 0$.}
    \label{Fig:Schematics_&_Fields}
\end{figure}

Here, we introduce the Dirac wire [see Fig. \subref*{Fig:Schematic}], the fermionic dual of the optical fiber. This system is the cylindrical generalization of the $m>0$, $m<0$ domain wall introduced by Jackiw and Rebbi \cite{jackiw1976solitons}; the canonical planar system which spurred the field of topological materials. Pioneering work has shown a null expectation value for the relativistic electron spin in the planar Jackiw-Rebbi problem \cite{bliokh2015quantum}. In stark contrast, we find that the Dirac wire supports longitudinal fermionic spin along its axis. For completeness, we also mention that the two-dimensional (2D) photonic dual of the Jackiw-Rebbi domain wall was discovered only recently \cite{gorlach2019photonic}, and is described by the interface of positive/negative gyrotropic media. Comparing Maxwell's equations to the 2D Dirac equation, the gyrotropic non-reciprocity coefficient was shown to play the role of photonic mass \cite{van2018quantum,VanMechelen:19,Horsley2018}.

The radius of the proposed Dirac wire is on the order of the Compton wavelength of the electron; fundamentally different from the well-known quantum wire limit \cite{datta1997electronic,latyshev2018manifestations}. We directly capture the relativistic effects of spin-orbit coupling and spin quantization in the spatial dynamics of the electron wavefunction. This allows us to explicitly show the half-integer quantization of the total angular momentum in an inhomogeneous waveguide system. This presents a unique approach to analyzing spin-orbit coupling in confined geometries, compared to traditional bulk energy band structure \cite{fu2007topological,hasan2010colloquium}. Solutions of the Dirac equation in a cylindrical geometry have been studied in the context of quantum chromodynamics \cite{jiang2016pairing}, however the relativistic spin-orbit coupling arising from confinement, spatial dynamics of spin, as well as the connection to the Jackiw-Rebbi problem have remained unexplored. Here, we analyze cylindrical generalizations of both the Jackiw-Rebbi domain wall and the minimal topological insulator \cite{shen2011topological,shen2012topological}, which will be of interest in spintronics and majorana physics \cite{Fan2016,larocque2018twisted,elliott2015colloquium}.


\textit{Dirac Wire.---}
We describe the Dirac wire as a cylinder with an effective electronic mass $m_1$, surrounded by a medium with an effective electronic mass $m_2$ [Fig.~\subref*{Fig:Schematic}]. The wire radius $a\approx \lambda_c$ is on the order of the Compton wavelength of the electron $\lambda_c = h/(m_1v_\text{F})$, where $h$, $m_1$, and $v_\text{F}$ are the Planck constant, electron mass, and Fermi velocity within the wire, respectively. We introduce three distinct classes of Jackiw-Rebbi (JR) domains labeled as JR$^+$, JR$^-$, and JR-D [Fig.~\subref*{Fig:Schematic}]. We also show important fundamental differences between cylindrical JR solutions (Dirac wires) and the conventional planar interface problem \cite{jackiw1976solitons} widely studied in the field of topological insulators and majorana physics \cite{elliott2015colloquium}. The main differences between the cylindrical and planar JR problems are the emergence of a longitudinal component of spin and the existence of confined solutions for all-positive electronic mass.


For a cylindrical Dirac waveguide, the difference in electronic mass inside and outside the wire gives rise to bound fermionic waves. These solutions can be derived from the time-independent Dirac equation,
\begin{equation}\label{Eq: Dirac Equation}
  H{\psi}_\mu= \left( v_\text{F} ~ \pmb{\alpha} \cdot \pmb{p} +  m v_\text{F}^2\beta\right){\psi}_\mu=E {\psi}_\mu.
\end{equation}
Eigenstates of the Dirac equation can be identified by five good quantum numbers which correspond to five commuting operators. In cylindrical coordinates, these operators are the Hamiltonian $H$, longitudinal total angular momentum $J_z$, longitudinal momentum $p_z$, transverse momentum $p_\bot^2$, and the transverse helicity $h_\bot$ \cite{balantekin1995second,jiang2016pairing}. The quantum numbers corresponding to these operators respectively are $E$, $\hbar\mu$, $\hbar k_z$, $\hbar k_\bot$, and $s=\pm 1$, where $\mu\in \mathbb{Z}+\frac{1}{2}$ is half-integer due to the fermionic nature of electrons. The two solutions corresponding to the two eigenvalues of transverse helicity $s=\pm 1$ are (see supp. info. \cite{supplementary}),
\begin{equation}\label{Eq:General_Eigenfucntions}
	\bm{u}^{(\pm)}_{\mu,M}(k)=\frac{C_\mu e^{ik_z z} e^{i\mu \phi}}{\sqrt{2}}
    \begin{pmatrix}
      Z_{n_+}(k_{\bot} \rho) e^{-i\phi/2} \\
      \pm Z_{n_-}(k_{\bot} \rho) e^{+i\phi/2}\\
      \mp i\hbar v_\text{F} \frac{k_{\bot} +ik_z}{M} Z_{n_+}(k_{\bot} \rho) e^{-i\phi/2} \\
      i \hbar v_\text{F} \frac{k_{\bot} +ik_z}{M} Z_{n_-}(k_{\bot} \rho) e^{+i\phi/2}
    \end{pmatrix}
\end{equation}
where $C_\mu$ is the normalization factor, $M = E+ mv_\text{F}^2$, $k_\bot=\sqrt{k^2-k_z^2}$, and $n_+-\frac{1}{2}=n_-+\frac{1}{2}=\mu$. Here, $\hbar^2k^2$ are the eigenvalues of total momentum operator $\pmb{p}^2$, and $n_{\pm}\in \mathbb{Z}$ are integers. The $s=\pm 1$ signs appearing in Eq.~(\ref{Eq:General_Eigenfucntions}) refer to the eigenvalues of the transverse helicity operator, $h_\bot$. $Z_n(k_\bot \rho)$ is a Bessel function of order $n$ and argument $k_\bot\rho$, where $\rho$ is the radial coordinate.


The vector spin operator of the Dirac equation is defined as $ \hat{\pmb{\Sigma}} = \frac{\hbar}{2}\begin{pmatrix}
                                    \pmb{\sigma} & 0  \\
                                    0     &  \pmb{\sigma}
                                    \end{pmatrix}
$,
where $\pmb{\sigma}=(\sigma_x,\sigma_y,\sigma_z)$ are the Pauli matrices expressed in vector operator form. The longitudinal component of the orbital angular momentum (OAM) operator is $
    \hat{L}_z = -i\hbar \frac{\partial}{\partial \phi} $.
Together with the spin operator, we obtain the longitudinal total angular momentum $\hat{J}_z = \hat{\Sigma}_z + \hat{L}_z$. In the subsequent sections we will use these operators to find the expectation values of the spin and orbital angular momentum of the modes.

\textit{Cylindrical Jackiw-Rebbi domain wall.---} We now solve the cylindrical wire geometry with an effective electronic mass $m_1$ surrounded by a medium with an effective electronic mass $m_2$. This is the cylindrical analogue of the 1D Jackiw-Rebbi (JR) domain wall \cite{shen2011topological,shen2012topological,jackiw1976solitons}. Unlike the 1D problem, however, solutions of the cylindrical geometry are not limited to the condition $m_1 m_2 <0$. Therefore, we analyze two separate cases; the case when $m_1,m_2>0$ and label it as JR$^+$, and the case when $m_1<0$, $m_2>0$ and label it as JR$^-$.

For the case of $m_1,m_2>0$ (JR$^+$), solutions of Eq.~(\ref{Eq: Dirac Equation}) only exist when $m_2>m_1$ which requires a larger mass (bandgap) outside the wire to confine the waves. This condition is analogous to total internal reflection in an optical fiber, which necessitates a lower refractive index outside the fiber \cite{okamoto2006fundamentals}. For the JR$^+$ problem, the solutions are characterized by $k_{\bot_1}$ real and $k_{\bot_2}$ imaginary where $k_{\bot_i}=\sqrt{k_i^2 - k_z^2}$ are the transverse (to the $z$-axis) propagation constants. $k_1$ and $k_2$ being the characteristic wavelengths inside and outside of the wire, respectively. Being comprised of evanescent waves outside the wire and standing waves inside, we denote these solutions as hybrid modes $H_{\mu,\nu}$. The subscripts $\mu$ and $\nu$ correspond to the total angular momentum eigenvalue and the order of the radial zero of the Bessel function.

\begin{figure}[t!]
    \centering
    \setlength\arrayrulewidth{1pt}
    \begin{tabular}{cc|c}
         & \Large $S_z$ & \Large $L_z$ \\
        \Large JR$^+$ &  \raisebox{-0.5\totalheight}{\label{Fig:Sz-01}\includegraphics[width = 0.19\textwidth]{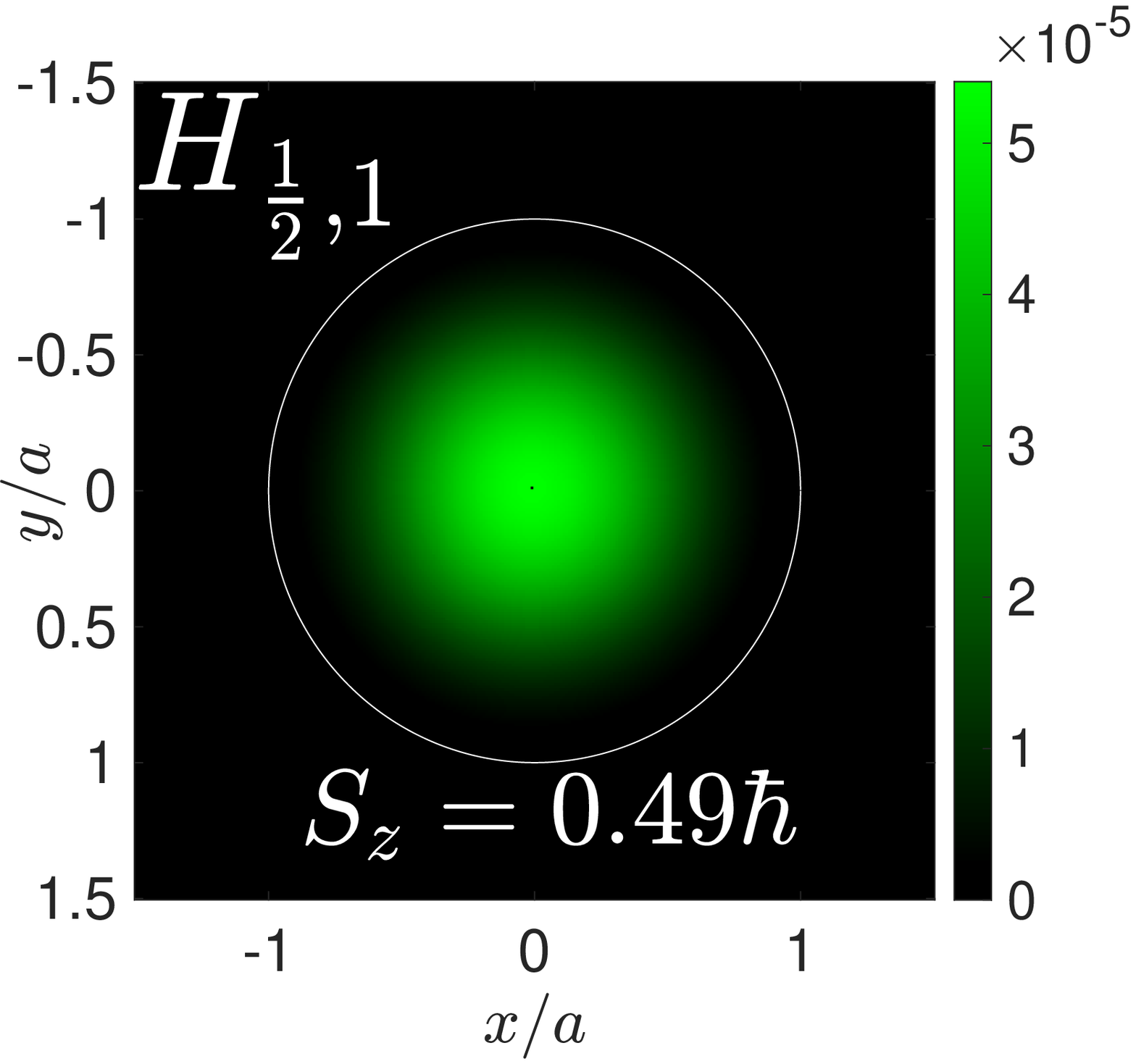}} & \raisebox{-0.5\totalheight}{\label{Fig:Lz-01}\includegraphics[width = 0.19\textwidth]{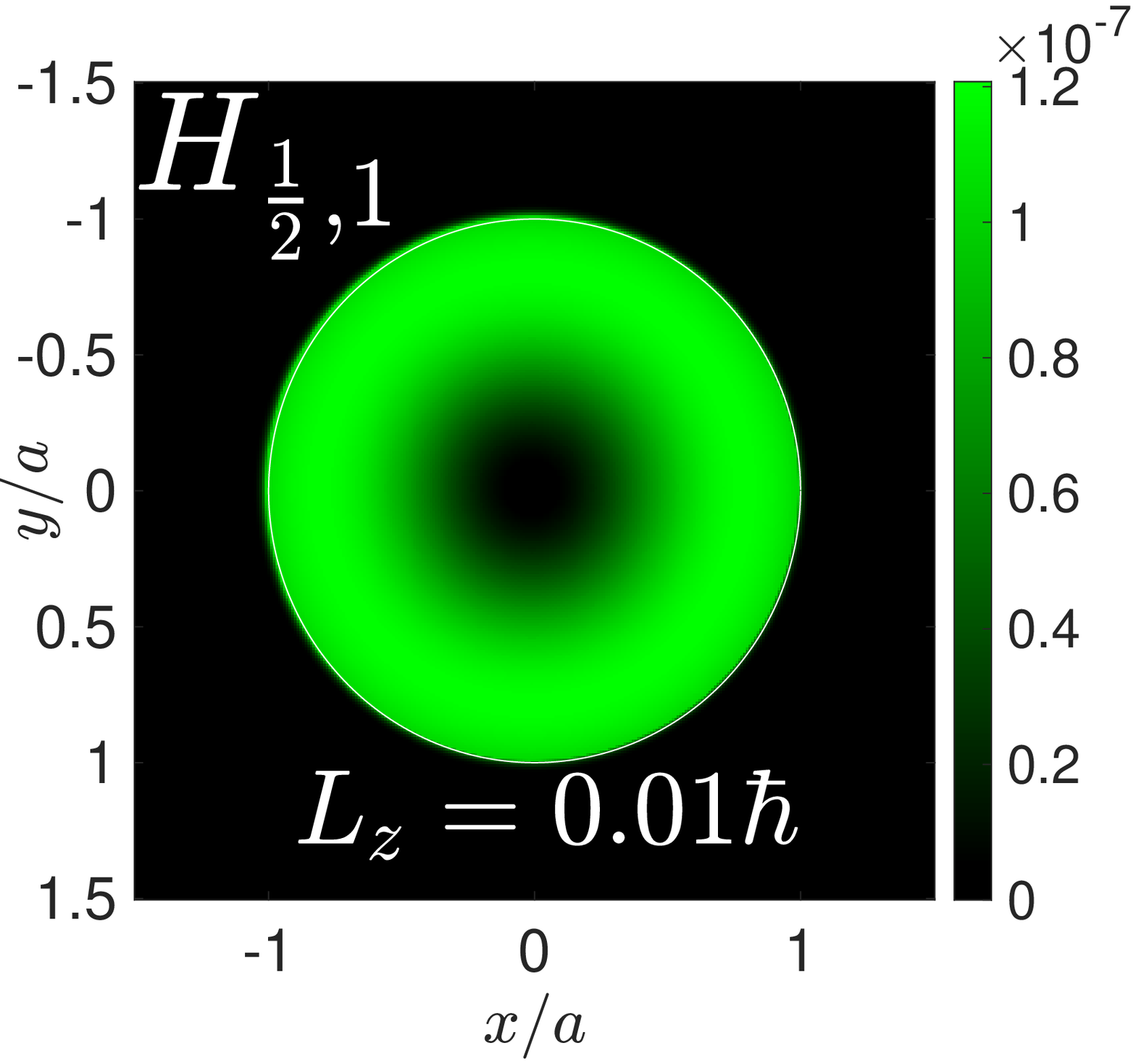}} \\
        \Large JR$^-$ &  \raisebox{-0.5\totalheight}{\label{Fig:Sz-02}\includegraphics[width = 0.19\textwidth]{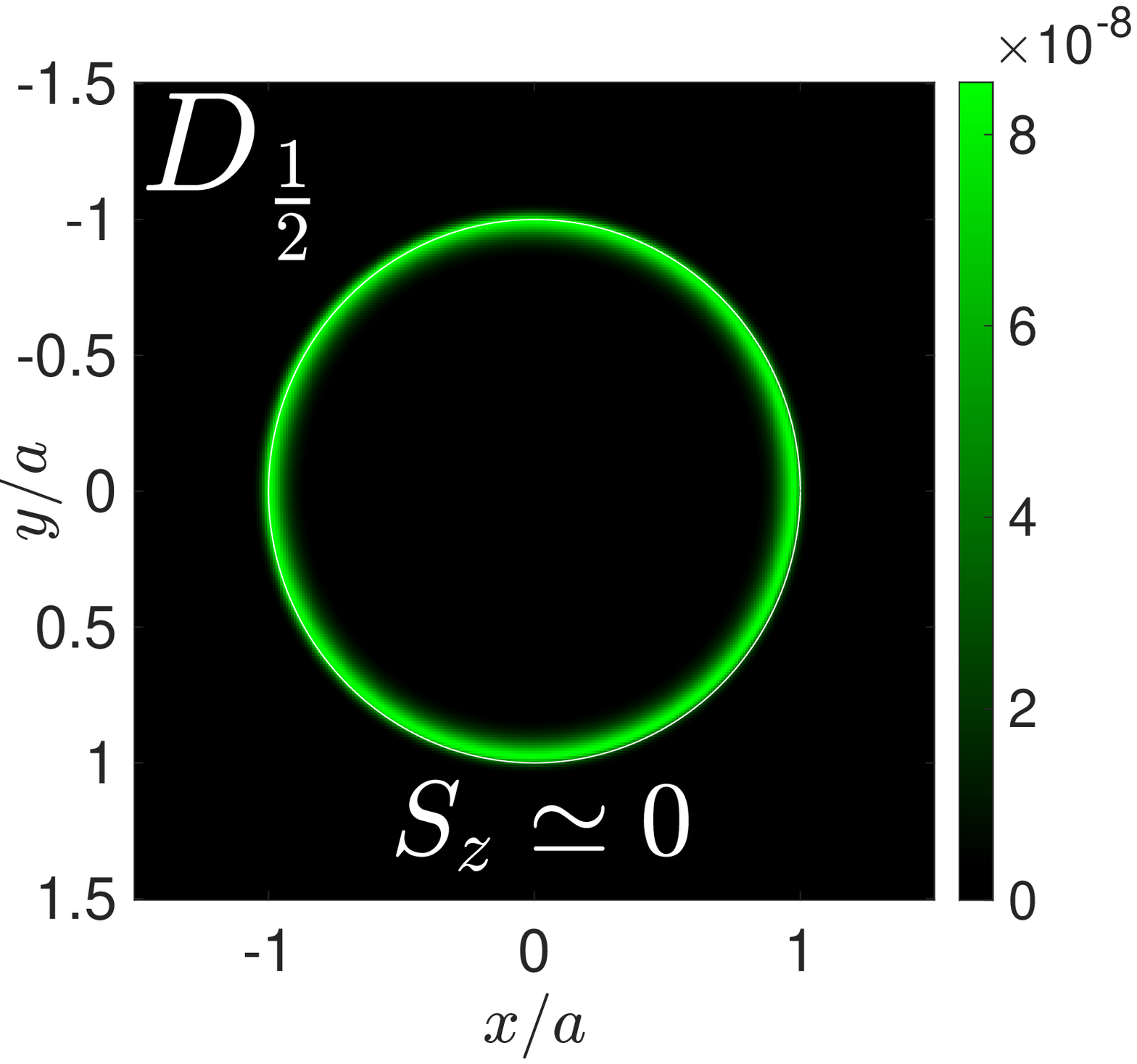}} & \raisebox{-0.5\totalheight}{\label{Fig:Lz-02}\includegraphics[width = 0.19\textwidth]{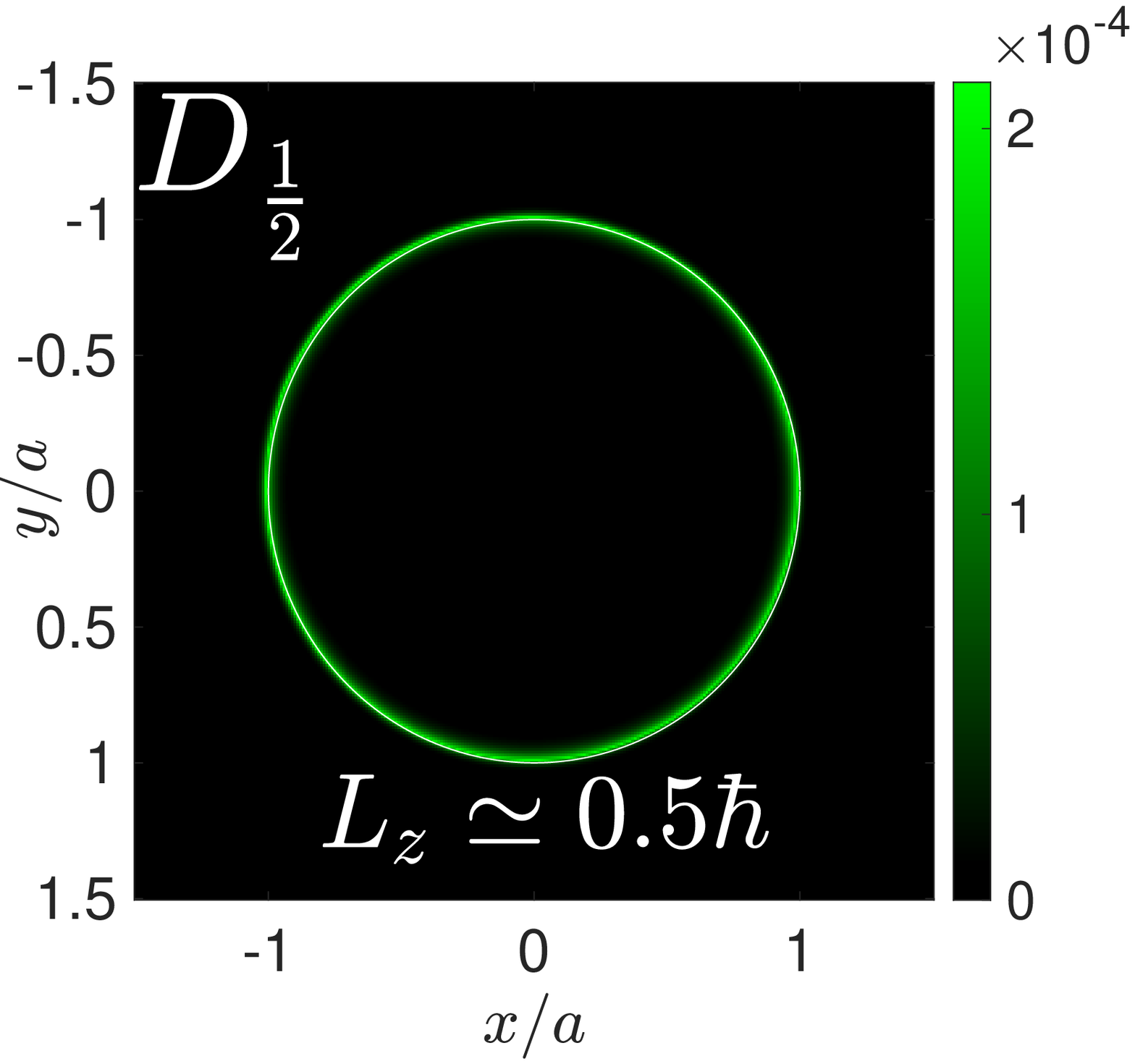}} \\
        \Large JR-D &
        \raisebox{-0.5\totalheight}{\label{Fig:Sz-03}\includegraphics[width = 0.19\textwidth]{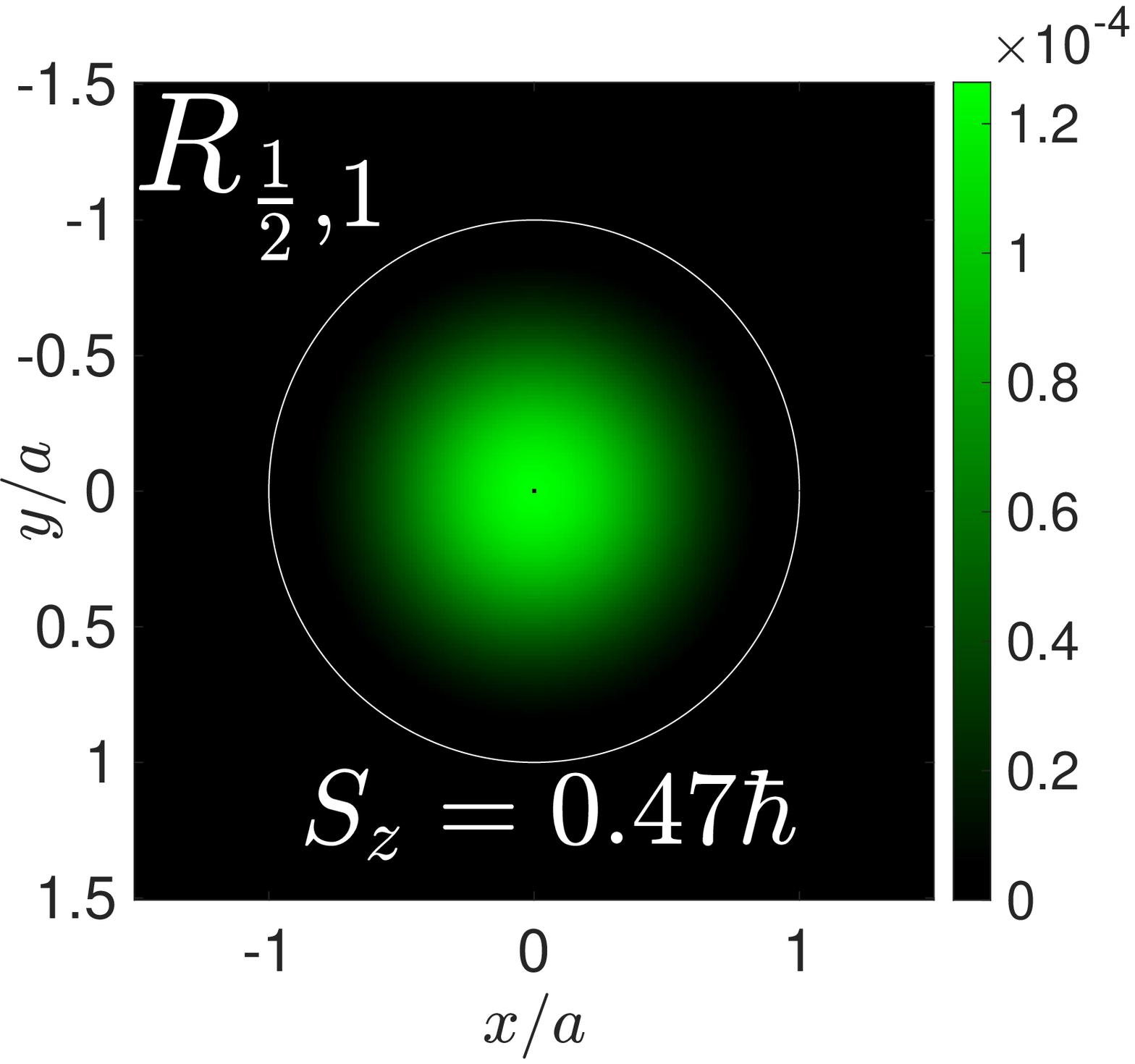}} & \raisebox{-0.5\totalheight}{\label{Fig:Lz-03}\includegraphics[width = 0.19\textwidth]{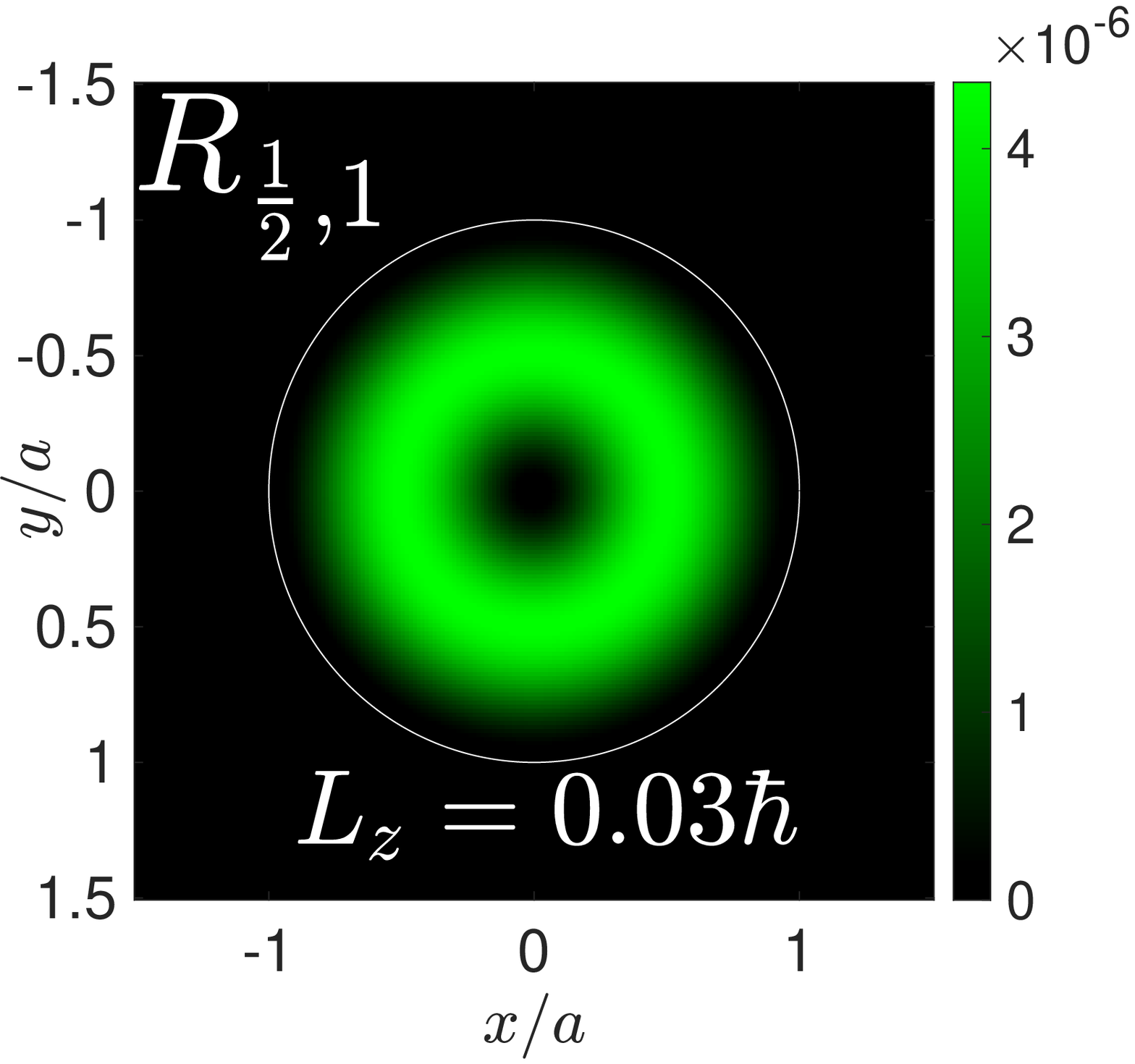}} \\
    \end{tabular}
    \caption{Spin and orbital angular momentum densities for the three Jackiw-Rebbi (JR) domains [Fig.~\protect\subref*{Fig:Schematic}]. As an example, $a=20$~\AA, $\mu= \frac{1}{2}$, and $k_z =0$. $|m_1|v_\text{F}^2$ and $m_2v_\text{F}^2$ are 1~eV and 2~eV, respectively. For the JR-D problem, $m_0v_\text{F}^2 = 1$~eV and $B\hbar^2 = 50$~eV\AA$^2$ \cite{zhang2009topological}. In all three scenarios we have assumed a Fermi velocity of $v_\text{F} \simeq 1.52\times10^5$~m/s, such that the Compton wavelength is $\lambda_c \simeq 8$~\AA. The values at the bottom of each figure are the integrated quantities of the respective distribution over the entire cross-section of the problem. Note that the spin and OAM are not individually conserved but their summation ($J_z = S_z + L_z$) is. Although not individually conserved, the difference in the distribution of spin and OAM makes them locally distinguishable. This means that one can, in principle, couple exclusively to spin or OAM locally.}
    \label{Fig:SzLz_JR}
\end{figure}
Figure \subref*{Fig:Field_profiles} (left panel) shows the amplitude of the wavefunction, $\psi^\dagger\psi$, for the dominant $H_{\frac{1}{2},1}$ mode. Note that for the JR$^+$ problem, the solutions vanish at $\rho\to\infty$ as the wavefunction is evanescent outside the wire. Figure \ref{Fig:SzLz_JR} (first row) displays the spatial distribution of longitudinal spin and orbital angular momentum densities for this mode. Note that the azimuthal $\hat{\phi}$ and radial $\hat{\rho}$ components of the spin and OAM are identically zero -- the angular momentum is purely longitudinal (directed along $\hat{z}$). The integrated values of spin and OAM over the entire $x-y$ plane is recorded at the bottom of each figure. For the $H_{\frac{1}{2},1}$ mode these values are not quantized, $S_z \simeq 0.49\hbar$ and $L_z \simeq 0.01\hbar$, respectively. Their sum, however, gives the half-integral value of $J_z = S_z + L_z = \frac{\hbar}{2} = \mu \hbar$ of the total angular momentum. These results show that while the spin and OAM are not separately conserved quantities, their sum, the total angular momentum, is conserved with an eigenvalue $\hbar\mu$. In other words, the wavefunctions ${\psi}_\mu$ are also eigenfunctions of the $\hat{J}_z$ operator \cite{bialynicki2017relativistic}.

Solutions for the Jackiw-Rebbi Dirac wire with $m_1<0$ and $m_2>0$ (JR$^-$) are similar to that of the JR$^+$ problem with the difference that, in addition to the hybrid $H_{\mu,\nu}$ modes, another set of solutions exists. These are characterized by decaying solutions outside and inside the wire ($k_{\bot_1}$ and $k_{\bot_2}$ both imaginary). We label these waves as decaying $D_{\mu}$ modes. In contrast to the hybrid modes, the decaying modes have only one possible solution for a given $\mu$ and are therefore labeled by only one quantum number (see supp. info. \cite{supplementary}). As shown in Fig.~\subref*{Fig:Field_profiles} (middle panel), the wavefunction of this mode is predominantly concentrated around the perimeter of the wire and is therefore the cylindrical analogue of the surface states in the planar Jackiw-Rebbi domain \cite{jackiw1976solitons}. In fact, as shown later, the gapless edge states of the planar geometry emerge when $a\to \infty$. The second row in Fig.~\ref{Fig:SzLz_JR} shows the spatial distribution of longitudinal spin and orbital angular momentum densities of the dominant mode, $D_{\frac{1}{2}}$, for the JR$^-$ problem. Here also, the spin and OAM are purely longitudinal due to the confinement. This is in stark contrast with the plane wave solutions of Dirac equation where the propagation direction of the electron does not put any constraint on the dirction of spin. In the Dirac wire, however, the direction of spin the electron is fixed by the axis of the wire. The integrated values of spin and OAM give $S_z\simeq 0$ and $L_z \simeq 0.5 \hbar$, respectively, which again produces $J_z = S_z + L_z = \frac{\hbar}{2}$.

\textit{Dispersive Jackiw-Rebbi (topological insulator).---}
We now solve the Dirac Hamiltonian in Eq.~(\ref{Eq: Dirac Equation}) when the electronic mass inside the wire is dispersive \cite{bernevig2013topological,shen2011topological},
\begin{equation}\label{Eq:Dispersive_mass}
    m_1v_\text{F}^2 = m_0v_\text{F}^2 - B\hbar^2 k ^2,
\end{equation}
where $m_0$ is the electron rest mass in the wire and $B$ is the dispersion factor. Denoted by JR-D, the dispersive mass gives rise to solutions satisfying open boundary conditions ($\psi_\mu = 0 $) on the surface of the wire, irrespective of the mass outside. This is confirmed by the plot of the probability density [right panel in Fig.~\subref*{Fig:Field_profiles}], where the wavefunction is identically zero for $\rho \geq a$. The dispersive mass considered here is the simplest model that produces the gapless edge states on the surface of a topological insulator \cite{liu2010model,bernevig2013topological,shen2011topological}. It can be shown that the bulk $\mathbb{Z}_2$ invariant is nontrivial $(-1)^\zeta = \mathrm{sgn}(-m_0B)$ whenever $m_0 B>0$. Hence, the medium $\rho<a$ is a topological insulator $\zeta=1$. Note, we do not consider the inverse problem in this paper, where the medium $\rho>a$ is topological and the wire is treated as a cylindrical defect.

\begin{figure}[t!]
        \centering
        \subfloat[\label{Fig:Dispersions}]{
        \includegraphics[width=0.47\linewidth]{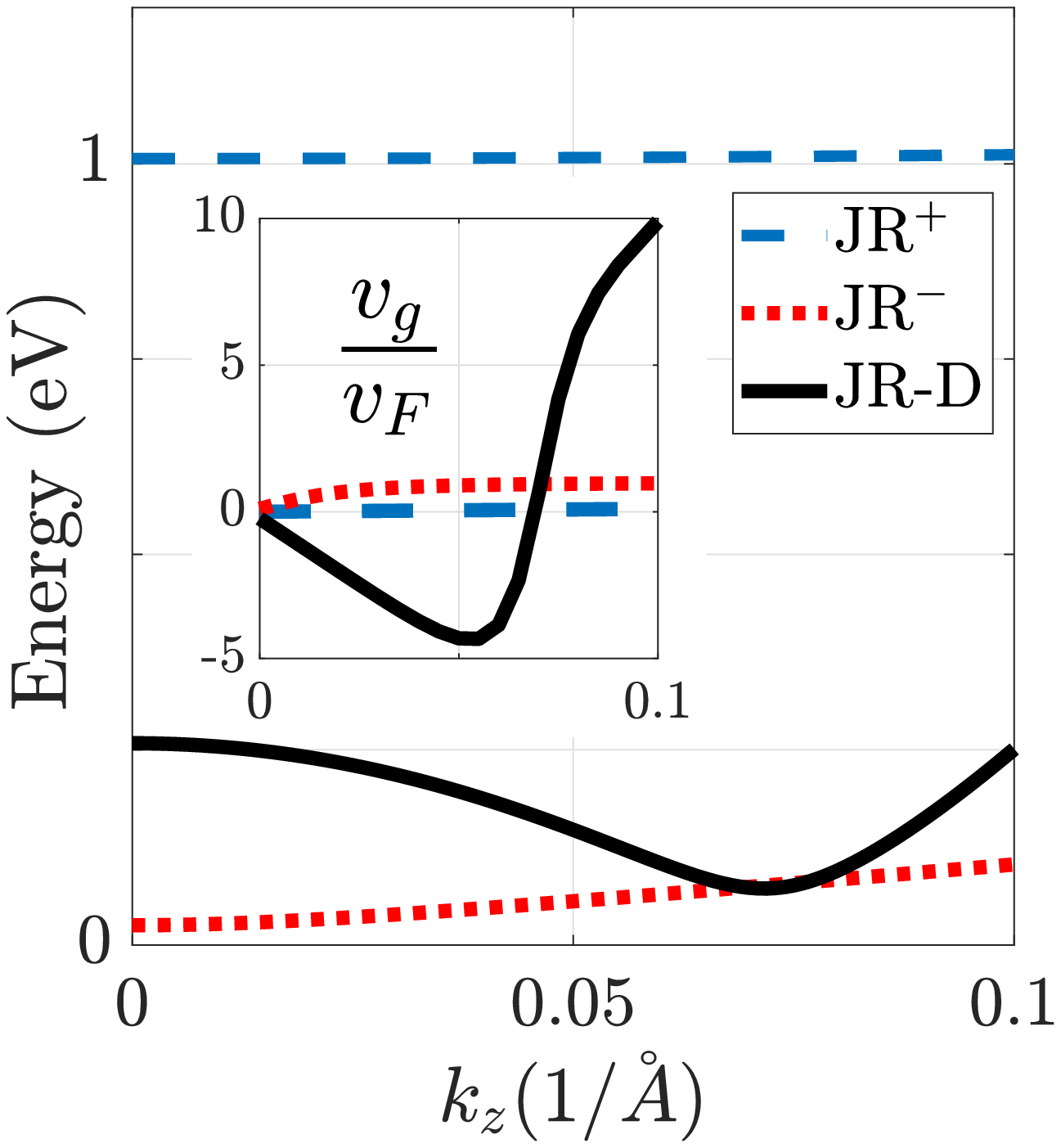}}
        \subfloat[\label{Fig:Bandgaps}]{
        \includegraphics[width=0.47\linewidth]{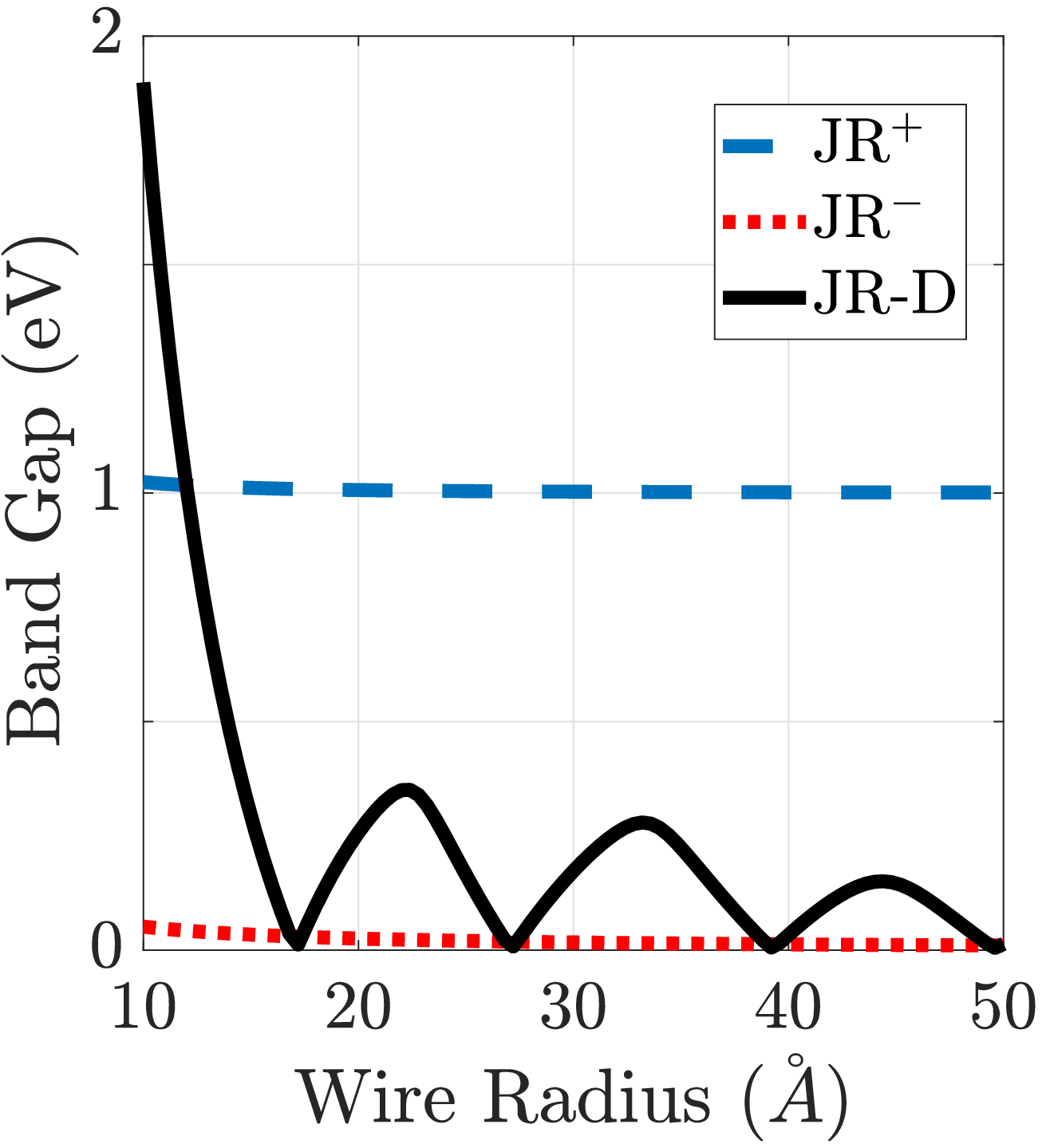}}\par \vspace{-4mm}
        \subfloat[\label{Fig:Spin_vs_Radius}]{
        \includegraphics[width=0.47\linewidth]{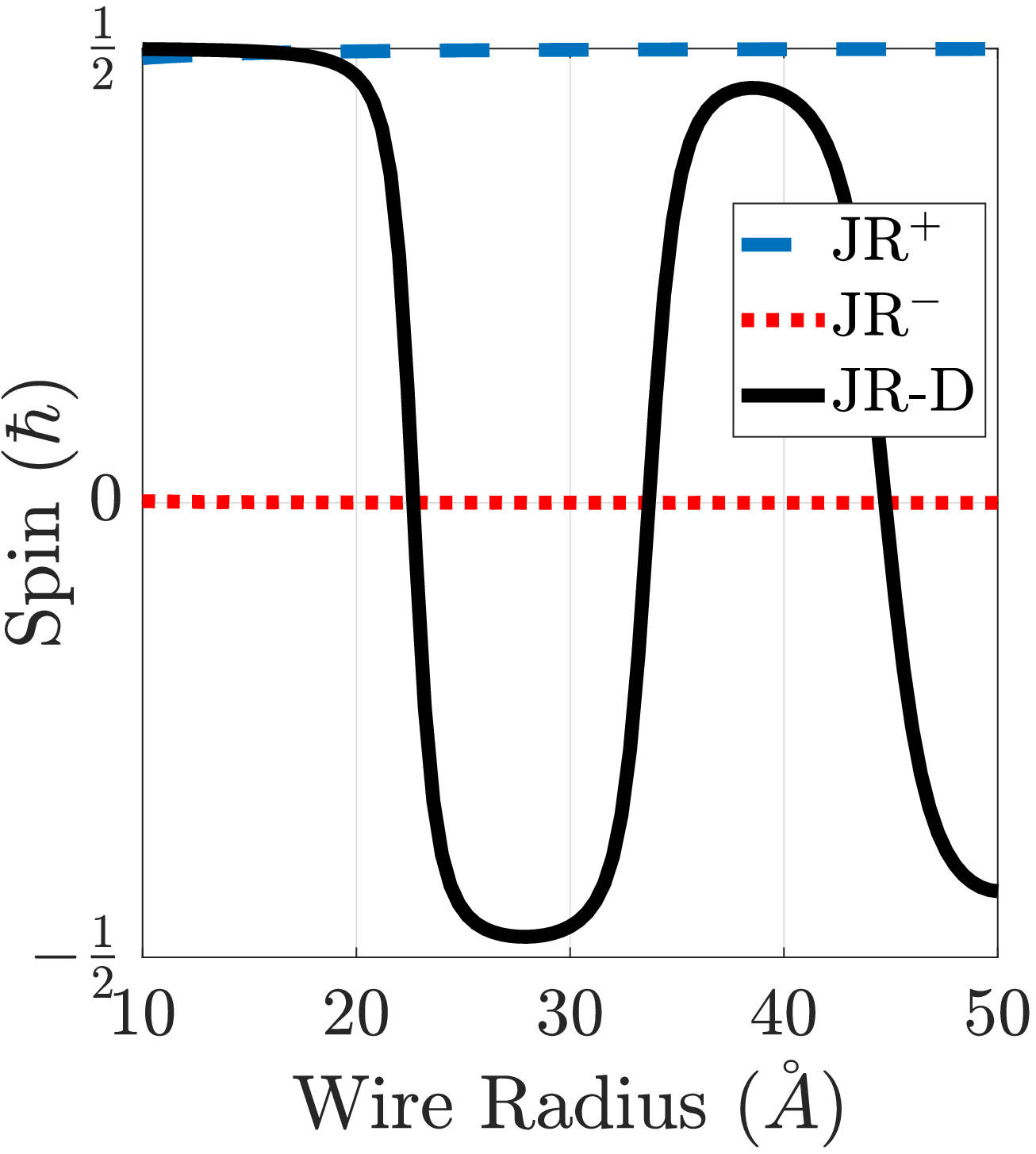}}
        \subfloat[\label{Fig:OAM_vs_Radius}]{
        \includegraphics[width=0.47\linewidth]{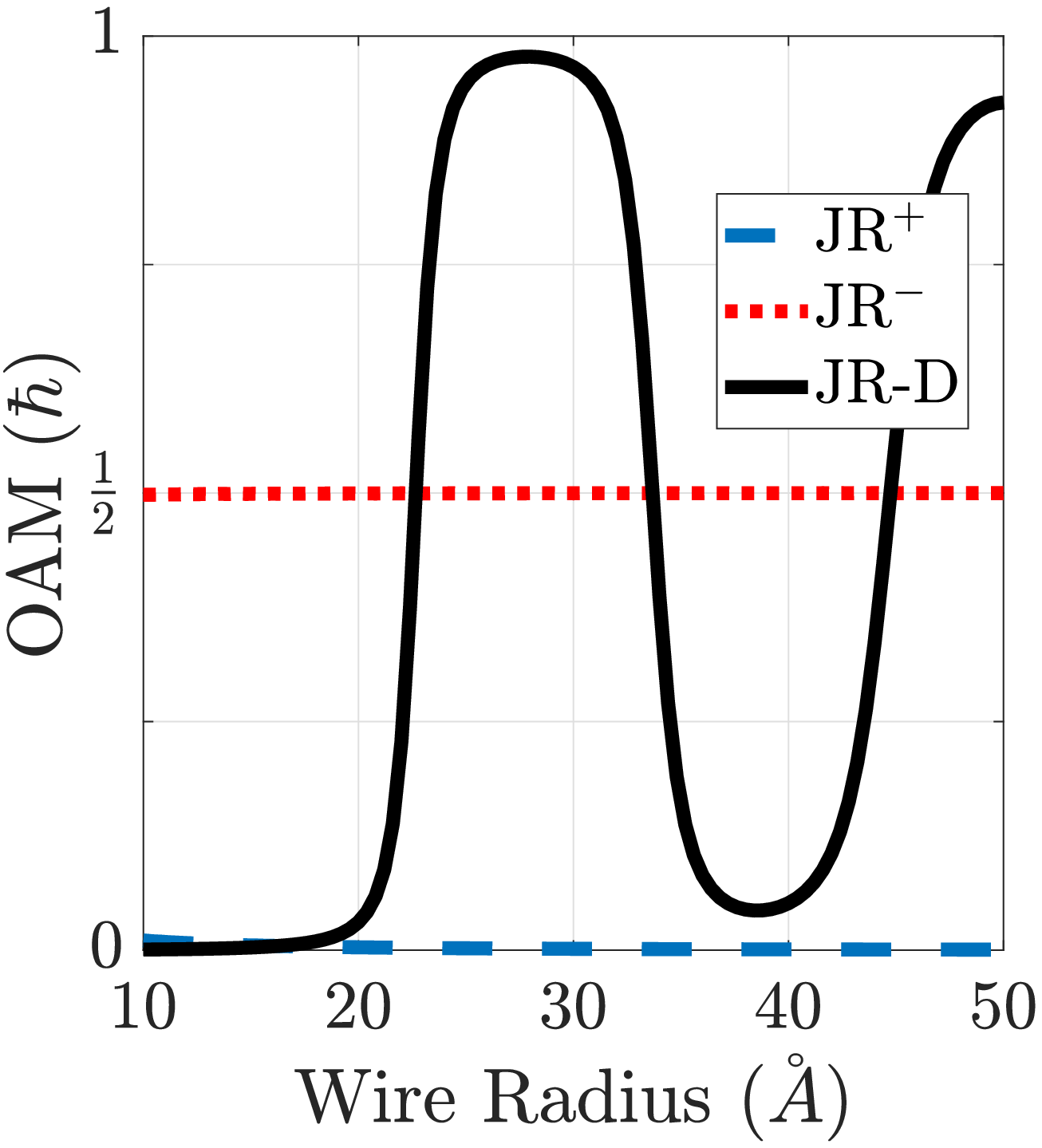}}
    \caption{(a) Dispersion and group velocities (inset) for the dominant modes of JR$^+$ (dashed blue), JR$^-$ (dotted red), and JR-D (solid black). Group velocities are normalized to the Fermi velocity $v_\text{F} \simeq 1.52\times10^5$ m/s. Wire radius dependence of (b) bandgaps, (c) spin, and (d) OAM for the three problems at $k_z = 0$. Here $\mu= \frac{1}{2}$, $k_z =0$, $|m_1|v_\text{F}^2 = 1$~eV, $m_2v_\text{F}^2=2$~eV, and $\lambda_c \simeq 8$~\AA. For the topological insulator (JR-D), $m_0v_\text{F}^2 = 1$~eV and $B\hbar^2 = 50$~eV\AA$^2$. Due to confinement in the cylindrical geometry, the bandgap is opened for all three problems. For JR-D, however, the bandgap closes and reopens for certain values of $a$ as seen in panel (b). Note that the summation $J_z = S_z + L_z$ produces the conserved value of $\frac{1}{2}$ in all three cases. In the limit $a\to \infty$, OAM vanishes $L_z \to 0$ for JR$^+$, while spin vanishes $S_z \to 0$ for JR$^-$ and JR-D.}
    \label{Fig:Radius_dependence}
\end{figure}

In the JR-D case, the eigenfunctions are of similar form as Eq.~(\ref{Eq:General_Eigenfucntions}) with the difference that instead of two, there are four eigenfunctions:
\begin{equation}\label{Eq:T-DW-Eigenfunctions}
    \begin{split}
    	\bm{u}^{(+)}_{\mu,M^{(1)}}(k^{(1)}),\qquad \bm{u}^{(-)}_{\mu,M^{(1)}}(k^{(1)}),\\ \vspace{4mm} \bm{u}^{(+)}_{\mu,M^{(2)}}(k^{(2)}),\qquad \bm{u^{(-)}}_{\mu,M^{(2)}}(k^{(2)}),
    \end{split}
\end{equation}
where $M^{(i)} = E + m_0 v^2 - B\hbar^2 (k^{(i)})^2$ and $\bm{u}^{(\pm)}_{\mu,M^{(i)}}(k^{(i)})$'s are given by Eq.~(\ref{Eq:General_Eigenfucntions}). Here $k_\bot^{(i)} = \sqrt{(k^{(i)})^2 - k_z^2}$ with $k^{(i)}$ being two possible propagation constants within the wire, resulting from the dispersive mass \cite{supplementary},
\begin{equation}\label{Eq:JRD_k}
  k^{(1,2)}=\frac{v_\text{F}}{\sqrt{2}B\hbar}\left[(2m_0B-1)\pm\sqrt{(1-4m_0B)+\frac{4B^2 E^2}{v_\text{F}^4}}\right]^\frac{1}{2}
\end{equation}
Unlike 1D solutions of the topological insulator \cite{shen2011topological,shen2012topological}, solutions of the cylindrical JR-D problem exist irrespective of the sign of $m_0B$. In this paper, however, we only consider the scenario when $m_0B>0$ since the solutions of the trivial case $\zeta = 0$ are similar to the JR$^+$ domain and are not particularly interesting.

Like the JR$^{\pm}$ states, we can label the modes depending on whether the two transverse propagation constants, $k_\bot^{(1)}$ and $k_\bot^{(2)}$ are real or imaginary. Note that $k_{\bot}^{(1)}$ and $k_{\bot}^{(2)}$ both belong to the interior of the wire $\rho< a$ as there are now two characteristic wavelengths [Eq.~(\ref{Eq:JRD_k})]. In addition to $H_{\mu,\nu}$ and $D_{\mu}$, two other types of modes labeled as $R_{\mu,\nu}$ and $C_{\mu,\nu}$ exist in the JR-D problem. These modes refer to real ($R_{\mu,\nu}$) and complex ($C_{\mu,\nu}$) solutions for $k_\bot^{(1,2)}$, respectively. The third row of Fig.~\ref{Fig:SzLz_JR} shows the spin and orbital angular momenta densities for the dominant mode, $R_{\frac{1}{2},1}$, of the JR-D problem. Here also, the azimuthal and radial components of the spin and OAM are identically zero -- only the longitudinal part is non-vanishing. Due to spin-orbit coupling, the spin and orbital angular momentum are not individually conserved. This means it is difficult to distinguish between the separate contributions of the total angular momentum in an experiment. The spatial distributions of spin and OAM in Fig.~\ref{Fig:SzLz_JR}, however, suggest a way to observe the spin or orbital parts locally. Analyzing the spin and orbital parts of $R_{\frac{1}{2},1}$ for the JR-D problem, for instance, we observe that while the spin is dominantly at the center of the wire, the orbital angular momentum is zero here and is distributed closer to the perimeter. This shows that using a point contact at the center of the wire, one can exclusively couple to the local spin of the $R_{\frac{1}{2},1}$ mode where the orbital angular momentum vanishes.

The dispersion relation $E=E(k_z)$ of the dominant modes is presented in Fig.~\subref*{Fig:Dispersions} and shows significantly larger group velocities for JR$^-$ and JR-D compared to JR$^+$, which implies higher conductivity. Anomalous dispersion for JR-D can be explained by the fact that, due to the dispersive electronic mass, charge transport is dominated by holes rather than electrons. This means that, in the regions where the group velocity becomes negative [inset of Fig.~\subref*{Fig:Dispersions}], charge currents propagate along the negative $\hat{z}$ direction for $k_z >0$ \cite{supplementary}. In the JR$^+$ problem, the bandgap plateaus to $m_1$ (1 eV) for large radii [Fig.~\subref*{Fig:Bandgaps}]. Since the $H_{\mu,\nu}$ modes of the JR$^{+}$ domain wall are mostly distributed within the bulk of the wire [Fig.~\subref*{Fig:Field_profiles}], these modes transform into bulk states of the Dirac equation when $a \to \infty$. For JR$^-$, on the other hand, the bandgap closes when $a\to \infty$. This can be explained by the fact that the mode is predominantly distributed around the perimeter of the wire [middle panel of Fig.~\subref*{Fig:Field_profiles}]. Therefore, the $D_{\frac{1}{2}}$ mode transforms into the edge states of the conventional 1D Jackiw-Rebbi problem \cite{jackiw1976solitons} when $a \to \infty$. The opening of the band-gap in the JR$^-$ problem, for small wire radius, can be explained by the hybridization of the edge state modes \cite{lu2010massive}.

More interesting is the bandgap of the topological insulator (JR-D) where for some finite values of radii, the bandgap closes and re-opens in an oscillatory fashion with $a$. For JR-D, spin also exhibits oscillatory behavior and passes through regions of positive and negative $S_z$ upon increasing the wire radius [Fig.~\subref*{Fig:Spin_vs_Radius}]. However, as $a\to \infty$, angular momentum is dominated by spin for the JR$^+$ problem and conversely dominated by OAM for JR$^-$ and JR-D. This means the dominant JR$^-$ and JR-D modes behave like edge states in the limit $a\to \infty$ and circulate around the perimeter of the material. Another important observation in Fig.~\subref*{Fig:Spin_vs_Radius} is that, although the spin is not conserved in any problem, its absolute value never exceeds $\frac{1}{2}$. This holds for all higher orders of $\mu$ and $\nu$ as well \cite{supplementary}. Note also, for all three cases, the total angular momentum is still conserved irrespective of the value of the wire radius.

\textit{Conclusion.}---
Our results show important differences between the 1D JR \cite{jackiw1976solitons,shen2011topological} and the cylindrical JR domain walls. In contrast to the 1D problem, the confined geometry of JR$^{\pm}$ and JR-D display non-zero longitudinal spin and orbital angular momentum. Moreover, we have shown that a sign change in mass is not necessary for the existence of confined cylindrical solutions of the Dirac equation. Labeled by JR$^+$, these Dirac waveguide solutions are the electronic analogue of the guided modes of an optical fiber \cite{okamoto2006fundamentals}. This observation makes wire geometry an excellent candidate as a Dirac waveguide, where electronic wave packets can propagate inside the wire with high confinement. While the experimental observation of these effects is challenging for a wire of this radius, we believe our results will push current techniques further due to their importance in spintronics and electron transport. Topological insulator nanowires of radius $a=20$~nm have been reported in the literature \cite{tian2013dual}. Although the JR-D problem has the simplest model for topological insulators, the parameters used here are within the range of real materials. For Bi$_2$Se$_3$, for instance, a Fermi velocity of $v_\text{F} \simeq 5.0 \times 10^5 m/s$, a band-gap of about $0.28$~eV, and also dispersion factor of $56.6$~eV\AA$^2$ has been reported \cite{zhang2009topological}. The parameters for other topological insulators such as Bi$_2$Te$_3$ and Sb$_2$Te$_3$ show that the Dirac wire is realizable using available materials \cite{liu2010model}.


\textit{Acknowledgements.---}
This research was supported by Alberta Innovates Technology Future (AITF) Scholarship as well as the DARPA Nascent Light-Matter Interactions program.

\bibliographystyle{apsrev4-2}
\bibliography{apssamp}
\end{document}